\documentclass[prl,aps,twocolumn,twoside,superscriptaddress]{revtex4}
\usepackage[T1]{fontenc}
\usepackage[latin2]{inputenc}

\usepackage{amsmath}
\usepackage{latexsym}
\usepackage{bbm}

\usepackage{color}
\begin{document}
\def\Real{{\hbox{\Bbb R}}} \def\C{{\hbox {\Bbb C}}}
\def\spec{R_{\alpha\beta}}

\newcommand{\ket}[1]{| #1 \rangle}
\newcommand{\bra}[1]{\langle #1 |}
\newcommand{\proj}[1]{| #1 \rangle   \langle #1 |}

\def\eig{|\lambda_\svec;\lambda_\pvec;\lambda_q\>}
\def\svec{{\bold{s}}}
\def\pvec{{\bold{p}}}

\def\kvec{{\bbox{k}}}
\def\lvec{{\bbox{l}}}
\def\duzomniejsze{<\kern-.7mm<}
\def\duzowieksze{>\kern-.7mm>}
\def\intlarge{\mathop{\int}\limits}
\def\textbf#1{{\bf #1}}
\def\beq{\begin{equation}}
\def\eeq{\end{equation}}
\def\be{\begin{equation}}
\def\ee{\end{equation}}
\def\ben{\begin{eqnarray}}
\def\een{\end{eqnarray}}
\def\beqa{\begin{eqnarray}}
\def\eeqa{\end{eqnarray}}
\def\eea{\end{array}}
\def\bea{\begin{array}}
\newcommand{\bei}{\begin{itemize}}
\newcommand{\eei}{\end{itemize}}
\newcommand{\bee}{\begin{enumerate}}
\newcommand{\eee}{\end{enumerate}}
\def\esr{SR}

\def\hcal{{\cal H}}
\def\pcal{{\cal P}}
\def\lcal{{\cal L}}
\def\acal{{\cal A}}
\def\ncal{{\cal N}}
\def\ecal{{\cal E}}
\def\dcal{{\cal D}}
\def\ccal{{\cal C}}
\def\ucal{{\cal U}}
\def\bcal{{\cal B}}
\def\rcal{{\cal R}}
\def\1{\openone}

\def\tr{{\rm Tr}}
\def\id{{\operatorname{id}}}
\def\ra{\rangle}
\def\la{\langle}
\def\>{\rangle}
\def\<{\langle}
\def\blacksquare{\vrule height 4pt width 3pt depth2pt}
\def\ic{I_{coh}}
\def\ot{\otimes}
\def\rhoab{\varrho_{AB}}
\def\sigmaab{\sigma_{AB}}
\def\rhoa{\varrho_{A}}
\def\sigmaa{\sigma_{A}}
\def\rhob{\varrho_{B}}
\def\sigmab{\sigma_{B}}
\def\dt#1{{{\kern -.0mm\rm d}}#1\,}
\def\squareforqed{\hbox{\rlap{$\sqcap$}$\sqcup$}}
\def\qed{\ifmmode\squareforqed\else{\unskip\nobreak\hfil
\penalty50\hskip1em\null\nobreak\hfil\squareforqed
\parfillskip=0pt\finalhyphendemerits=0\endgraf}\fi}

%%%%%%%%%%%%%%%%%%%%%%%%%%%%%%%%%%
\def\qp{I_{\text{ind}}}
\def\Q{I_{\text{ind}}}
\def\Qzero{I_{\text{ind}}}
\def\qpriv{{mutual\ independence}}
\def\Qpriv{{Mutual\ independence}}
%%%%%%%%%
%%%%%%%%%%%%%%%%%%%%%%%%%

\def\sigmaa{\sigma_A}
\def\sigmaae{\sigma_{AE}}
\def\sigmae{\sigma_E}

\def\phitabe{\tilde \phi_{ABE}}
\def\psitabe{\tilde \psi_{ABE}}
\def\psit{{\tilde \psi}}

\def\rk{{\rm rk\,}}

\def\supp{{\rm supp\,}}

\def\ep{\epsilon}

\def\lav{\bigl\<}
\def\rav{\bigl\>}

\def\lavv{\biggl\<}
\def\ravv{\biggl\>}

\def\EPR{{\tt EPR}}
\def\GHZ{{\tt GHZ}}
\def\wobie{\rightleftharpoons }

\newtheorem{lemma}{Lemma}
\newtheorem{corrolary}[lemma]{Corrolary}
\newtheorem{theorem}[lemma]{Theorem}
\newtheorem{proposition}[lemma]{Proposition}
\newtheorem{definition}[lemma]{Definition}
\newtheorem{conjecture}[lemma]{Conjecture}
\newtheorem{fact}[lemma]{Fact}
\newtheorem{propositionn}{Proposition}

\def\bep{\begin{proposition}}
\def\eep{\end{proposition}}
\def\bel{\begin{lemma}}
\def\eel{\end{lemma}}

\def\bet{\begin{theorem}}
\def\eet{\end{theorem}}
\def\bed{\begin{definition}}
\def\eed{\end{definition}}
\def\bef{\begin{fact}}
\def\eef{\end{fact}}

\def\Z{Z}

%%%%%%%%%%% for sup section
\def\prop3{3}
\def\con4{4}
\def\conjnolock{5}
\def\sec1a{1A}
\def\eqlocal{(10)}
\def\eqjqp{12}
\def\eqqpriv-dimensionbound{5}

%\language\english

%preprint style
%\documentstyle[preprint,aps]{revtex}

%preprint style - single spaced
%\documentstyle[tighten,aps]{revtex}

\title{Quantum mutual independence}

\author{Micha\l{} Horodecki}
\affiliation{Institute of Theoretical Physics and Astrophysics, University of Gda\'nsk, 80-952 Gda\'nsk, Poland}

\author{Jonathan Oppenheim}
\affiliation{Department of Applied Mathematics and Theoretical Physics, University of Cambridge, Cambridge CB3 0WA, U.K.}

\author{Andreas Winter}
\affiliation{Department of Mathematics, University of Bristol, Bristol BS8 1TW, U.K.}
\affiliation{Centre for Quantum Technologies, National University of Singapore,
 2 Science Drive 3, Singapore 117542}

\date{14 October 2009}

\begin{abstract}
We introduce the concept of \emph{mutual independence} 
-- correlations shared between distant parties
which are independent of the environment. This notion is more general than the
standard idea of a secret key -- it is a fully quantum and more general 
form of privacy. The states 
which possess mutual independence also 
generalize the so called private states -- 
those that possess private key. 
We then show that the problem of distributed compression
of quantum information at distant sources can be solved 
in terms of mutual independence, if free entanglement between the
senders and the receiver is available. 
Namely, we obtain a formula for the sum of rates of qubits needed 
to transmit a distributed state between Alice and Bob to a decoder Charlie.
We also show that mutual independence is bounded from above 
by the relative entropy modulo a conjecture, 
saying that if after removal of a single qubit the state becomes product, 
its initial entanglement is bounded by 1. 
We suspect that mutual independence is a highly singular quantity, i.e.~that it
is positive only on a set of measure zero; furthermore, we believe
that its presence is seen on the single copy level. 
This appears to be born out in the classical case. 
\end{abstract}

\maketitle

%\textcolor{green}{%
%\begin{itemize}
%  \item {\tt Dimension bound on $ab\alpha\beta$?}
%  \item {\tt Page 4: convex?}
%  \item {\tt Classical analogue and Conclusions!?}
%  \item {\tt Evidence for the Conjectures?}
%\end{itemize}}

%\section{Introduction}

{\bf \Qpriv\ -- Definition.}
In the paradigm of quantum key distribution \cite{BB84,Ekert91}, the goal is to share a 
cryptographic key -- perfectly 
correlated strings of bits, which are secure, i.e. are not correlated with any third person. 
Although we obtain a secret key from a quantum state, the secret key itself is a classical
object.  It also must be perfectly correlated, which is a very particular distribution.
States that possess secret 
key  have been fully characterized in \cite{pptkey,keyhuge} -- 
they are so called {\it private bits (pbits)}.  

Here we consider a more general 
paradigm, and one which is fully quantum.  Alice and Bob are interested in  
obtaining a state which is secure, in that it is not correlated with any 
third person, but the correlations contained in the state
need not be perfect, nor the state necessarily classical. 
We quantify the correlations by the mutual information. 
Such secure correlations we shall call {\it \qpriv}.  It can be understood
as quantum privacy.  

A protocol for extracting \qpriv, we shall call any 
sequence of local operations  $\Lambda_A^{(n)}$ and $\Lambda_B^{(n)}$ such that 
the state 
\be
  \rho^{(n)}_{ABR} = \bigl(\Lambda_A^{(n)} \ot \Lambda_B^{(n)} \ot \id_R \bigr) \psi^{\ot n}_{ABR}
\ee 
is asymptotically product in the cut $AB:R$, where $\psi_{ABR}$ is the 
purification of $\rho_{AB}$, i.e. 
\be
  \bigl\| \rho^{(n)}_{ABR} - \rho^{(n)}_{AB}\ot \rho_R^{(n)} \bigr\|_1 \to 0 
\ee
for $n\to \infty$, and $\| \xi \|_1 = \tr |\xi|$ being the trace norm.
Using the Stinespring dilations of
$\Lambda_A^{(n)}$ and $\Lambda_B^{(n)}$, we can say it in different 
words. Namely, a protocol extracting common \qpriv\  
amounts to  decomposing the local systems into two subsystems $\alpha$ 
and $a$  and $\beta$ and $b$, such that state $\rho_{\alpha \beta R}$  is product with 
respect to the cut $\alpha \beta : R$. Now the \qpriv\ will be the maximal amount of mutual
information between $\alpha$ and $\beta$ per copy of the initial state. 
The choice of this correlation measure may appear arbitrary here
-- and at this point any functional monotonic under local operations
would be eligible. However, the mutual information will find its
motivation in the section on distributed compression below.

\begin{definition}
Given state $\rho_{AB}$, consider a protocol of extracting 
\qpriv\ $\pcal={\Lambda_n}$. Define the rate 
\be
  R(\pcal,\rho_{AB}) = \liminf_{n\rightarrow\infty} \frac{1}{n} 
                \frac{1}{2}I\bigl( (\Lambda_A^{(n)} \ot \Lambda_B^{(n)}) \rho_{AB}^{\ot n} \bigr).
\ee 
Then \qpriv\ of state $\rho_{AB}$ is defined as 
\be
  \label{def:qpriv1}
  \qp(\rho_{AB}) = \sup_{\pcal} R(\pcal,\rho_{AB}).
\ee
For technical reasons we will only consider protocols such that there
exists a constant $c$ which (ratewise) bounds the output dimensions
of $\Lambda_A^{(n)}$ and $\Lambda_B^{(n)}$:
$|\alpha|,\,|\beta| \leq c^n$. Note that this implies similarly
$|a| \leq c^n |A|^n$ and $|b| \leq c^n |B|^n$, so that we can choose
a constant $r$ with
\be
  |a|, |b|, |\alpha|, |\beta| \leq r^n.
  \label{eq:qpriv-dimensionbound}
\ee
%
%\textcolor{red}{{\tt Any way that can be justified from the %definition?}}
%
\end{definition}

\medskip\noindent
{\bf Remark.} In a similar way one defines LOCC \qpriv\
$\qp^{\leftrightarrow}$, and one-way distillable 
\qpriv\ $\qp^{\rightarrow}$, where instead of local
operations $\Lambda_A^{(n)} \ot \Lambda_B^{(n)}$, more
general $\Lambda_{AB}^{(n)}$ implementable by LOCC and
one-way LOCC, respectively.  Let us also mention, 
that in \cite{buscemi-shredding}, another scenario involving 
decoupling a subsystem from reference was considered: 
there only one system  (e.g. $B$) was available, and the task was to split it into two parts.
%, and anlyse the tradeoff between mutual information of those parts with $R$.

Given these definitions, it would be good to know,
which states already have \qpriv. 

\begin{definition}
\label{defi:qpriv}
We say that the state $\rho_{ABA'B'}$
\emph{has \qpriv\ in $AB$} if two conditions are satisfied:
\begin{enumerate}
\item The state $\rho_{RAB}$  is product with respect to the cut $R:AB$,
    where $\rho_{RAB}=\tr_{A'B'} \psi_{RABA'B'}$, and
    $\psi$ is a purification of $\rho_{ABA'B'}$.
\item The state $\rho_{AB}$ is correlated. 
\end{enumerate}
\end{definition}
According to the definition of \qpriv{} in eq.~(\ref{def:qpriv1}), such states 
have therefore at least $\frac{1}{2} I(A:B)$ bits of \qpriv.
Note the factor of $1/2$, which we introduce for the sake of
normalisation: in this way, an ebit has one unit of \qpriv.

\medskip\noindent
{\it Multipartite case.}
For more than two parties, we define \qpriv\ in an analogous way.
To quantify it, we employ the following multipartite 
generalization of mutual information (sometimes called
\emph{multi-information}):
\be
  I(A_1: \ldots :A_N) = S(A_1) + \ldots + S(A_N) - S(A_1\ldots A_N).
\ee
We shall, however, mostly formulate our results for the bipartite
 case, and only occasionally hint at the $N$-party generalization.

States which posses mutual independence are a generalisation of pbits and for
two parties can be characterised by
\begin{proposition}
\label{prop:uhlmann}
A state $\rho_{ABA'B'}$ has \qpriv\ in systems 
$AB$, if and only if there exists an isometry $U:{A'B'} \rightarrow {CD}$ 
such that  
\be
  (\1_{AB} \ot U_{A'B'}) \rho_{ABA'B'} (\1_{AB} \ot U_{A'B'})^\dagger
     = \psi_{ABC}\ot \rho_{D},
\ee
with a pure state $\psi$ on $ABC$.
\end{proposition}
The proof of this, along with a discussion on the connection between pbits and the current work can be found in the Supplementary Materials.

\medskip
%{\bf Examples.}
%\label{subsec:maxcor}
Now, we want to focus on a few examples of \qpriv. As noted above, 
each ebit contains at least $1$ bit of \qpriv, each pbit has at least
$1/2$ bit of \qpriv. However, sometimes this might be an underestimate, as
one can continuously interpolate between a pbit and an ebit. 
In both cases, on the other hand, \qpriv\ is seen on the
level of a single copy of the state, and local collective 
operations will not obviously increase it. 

In the following example we show that \qpriv\ may require 
collective actions on many copies.

\medskip\noindent
{\bf Example.} The \qpriv\ of a maximally correlated state,
\be
  \rho_{AB} = \sum_{ij}a_{ij} |ii\>\<jj|,
\ee
is bounded from below by one-half of the \emph{coherent information}
$I(A\rangle B) = S(B)-S(AB) = S(A|R)$:
\be
  \qp(\rho_{AB}) \geq \frac{1}{2} I(A\rangle B).
\ee
{\bf Proof.}
We will simply show that by local unitaries, 
Alice and Bob can extract $S(B)-S(AB)$ amount of private key. 
To this end, assume that Alice and Bob 
share many copies of the state $\rho_{AB}$ and 
dephase their systems in the computational basis
(by copying them onto local ancillas).
Then they share many copies of a classical distribution with perfect correlations. 
To obtain the key it is therefore enough to apply hashing, which does not require communication. 
The amount of key is $I(A:B)-I(A:R)$,
which computed on the so called ccq-state obtained after dephasing 
gives $S(B)-S(AB)$ (the ``ccq'' means that Alice and Bob have a
classical register, while $R$ has a quantum register). \qed

The example of maximally correlated state suggests 
a wider class of states, which have \qpriv{},
but which require collective actions to distill it;
these are discussed in the Supplementary Materials.

%and the above states can be analyzed as they are in that framework.

\medskip\noindent
{\bf Possible discontinuity of \qpriv{}?}
Note that even though the maximally correlated state 
does not have \qpriv\ on the single copy level, we see some singularity 
in its structure, namely, it can be obtained 
by acting on a singlet with noise, whose errors 
do not span the full algebra: namely, there are no bit-flip errors.
In other words, the state comes from a channel, 
with a noiseless (classical) subsystem. 
E.g. we expect, that a Bell mixture of rank $3$ 
or $4$,  with fidelity arbitrary close to one, will have $\qp=0$,
even though $\qp=1$ for fidelity equal to one. 

Similarly, for the rank-two Bell mixture
$\rho = (1-\ep)\Phi^+ + \ep\Phi^-$, where the above example gives
a lower bound on the \qpriv\ of $\frac{1}{2}\bigl(1-H_2(\ep)\bigr)$:
we expect that for $0 < \ep < 1$ the \qpriv\ is $\qp \leq 1/2$, while
at $\ep=0$ and $1$ it is $\qp=1$.

\medskip
In an attempt at formalizing the above, we formulate the following
conjecture, which identifies the
presence of a ``noiseless (or rather private) subsystem'' 
in the correlations of $\rho_{AB}$. 

\begin{conjecture}
  \label{conj:algebra}
  A state $\rho_{AB}$ has $\qp(\rho_{AB}) > 0$ only if
  there exist operators $A$, $B$ not proportional to $\1$, such that
  for all states $\ket{\psi}$ in the support of $\rho_{AB}$,
  \begin{equation}
    \bra{\psi} A\otimes B \ket{\psi} = \tr\rho(A\otimes B).
  \end{equation}
\end{conjecture}
If the conjecture were true, it would mean that \qpriv\ is a singular
quantity: it could be positive only on a set of states of measure zero.
We discuss the conjecture in more detail in the Supplementary Materials. 
There, we also discuss the classical analogue to mutual independence 
and distributed compression, and in this context discuss the conjectured 
discontinuity which appears in a very simple form.

\medskip\noindent
{\bf Upper bounds.}
In~\cite[Theorem X.2]{AbeyesingheDHW2006-fullSW} one can find implicitly a proof
of the bound $\qp(\rho_{AB}) \leq E_{\text{sq}}(\rho_{AB})$,
the squashed entanglement~\cite{Winter-squashed-ent}. We explain it
in the Supplementary Materials.

Here we describe an attempt to upper bound the \qpriv\ 
by the relative entropy of entanglement~\cite{VPRK1997,PlenioVedral1998}
\be
  E_r(\rho)=\min_\sigma S(\rho\|\sigma),
\ee
where the minimum is taken over all separable states $\sigma$, and 
$S(\rho\|\sigma)=\tr \rho (\log \rho - \log \sigma)$ is the relative entropy.
Note that the relative entropy measure can be much smaller
than squashed entanglement~\cite{ChristandlW-lock}.
%\textcolor{red}{
This upper-bound hinges on the following conjecture:
\begin{conjecture}
  Consider a tripartite state $\rho_{XAB}$ such that $\rho_{AB}=\rho_A \ot \rho_B$. 
%  Then there exists a decomposition $\rho_{XAB}=\sum_i p_i %|\psi_i\>\<\psi_i|_{XAB}$ 
%  such that the pure states $\ket{\psi_i}$ have Schmidt rank at
% most
%  $|X|$ (the dimension of $X$) across the $XA:B$ cut.  
%Then convex roof of Schmidt rank (denoted later as $\esr$) 
Then the logarithmic negativity $E_N$ across the $XA:B$ cut  
does not exceed $\log |X|$. 
The logarithmic negativity is an entanglement  measure 
\cite{ZyczkowskiHSP-vol,Vidal-Werner} given by 
\be
  E_N(\rho) = \log \| \rho^\Gamma \|_1,
  \label{eq:logneg}
\ee
where $\Gamma = \id \ot \top$ is the partial transpose.
  \label{conj:nolock}
\end{conjecture}
%}
%Recall that convex roof $\hat f$ of a function $f$ is given by 
%\cite{Uhlmann-roof}
%\be
%\hat f(\rho)=\inf \sum_i p_i 
%f(|\psi_i\>\<\psi_i|)
%\ee
%where infimum is taken over all decompositions of state $\rho$
%into pure states.
The conjecture states, in other words, 
that if the initial state is product, then by providing $n$ qubits 
to one of the parties, entanglement can be increased at most by $n$.
%(if we think about logarithm of $\esr$). 
This should be compared with the locking effect \cite{lock-ent}: 
there, by adding one qubit, entanglement may be 
increased by an arbitrary amount.  However, in all the known examples 
for $E_N$, the initial state is non-product. 

We shall first consider states which have exact \qpriv\ in systems $AB$.
\begin{proposition}
Suppose that state $\rho_{ABA'B'}$ has \qpriv\ in systems $AB$. 
Then, assuming that Conjecture~\ref{conj:nolock} holds, we have 
\be
  E_r^\infty(\rho_{AA':BB'}) \geq \frac{1}{2} I(A:B),
\ee
where 
$E_r^\infty(\rho) = \lim_{n\rightarrow\infty} \frac{1}{n} E_r(\rho^{\ot n}) \leq E_r(\rho)$.
\end{proposition}
The proof goes via standard arguments using monotonicity and asymptotic continuity 
of relative entropy of entanglement, see e.g.~\cite{Michal2001}; its
details can be found in the Supplementary Materials.
Note that here we have to invoke again the technical dimension
condition in eq.~(\ref{eq:qpriv-dimensionbound}).

Because relative entropy of entanglement is monotonic under local
operations (indeed LOCC), the proposition implies that
$E_r^\infty(\rho_{AB}) \geq \qp^{\leftrightarrow}(\rho_{AB}) \geq \qp(\rho_{AB})$.

Of course, this means that also $E_r$ is an upper bound on \qpriv, 
since $E_r\geq E_r^\infty$. \qed
%The proof is given in the Supplementary Materials section.

\medskip\noindent
{\bf Distributed compression with free entanglement.}
%\label{sec:distributed}
The task of distributed compression is the following:
Alice and Bob share state  $\rho_{AB}^{\ot n}$ and let $\psi_{ABR}$ 
be a purification of $\rho_{AB}$. The goal is 
that some distant decoder  $C$ will share a state 
which will approach $\psi_{ABR}^{\ot n}$ for large $n$.
To this end Alice and Bob will independently send qubits to Charlie.
The problem is to find the region of pairs of rates $R_A$ and $R_B$
of sending qubits by Alice and Bob, respectively, to Charlie which achieve the goal. 
We assume that Charlie shares auxiliary entanglement 
with Alice and Bob, separately. Bounds to this region, without the
auxiliary entanglement, were given in~\cite{AbeyesingheDHW2006-fullSW}.
In particular it was shown that the rate pairs
\begin{align*}
  R_A &= \frac12 I(A:R),\ R_B = S(B), \\
  R_A &= S(A),\ R_B = \frac12 I(B:R),
\end{align*}
are achievable. Note that this implies the rate sum
$R_A+R_B = \frac{1}{2}J(A:B)$, with $J(A:B) := S(A)+S(B)+S(AB)$.
On the other hand, the rate region is bounded as follows:
\begin{align*}
  R_A     &\geq \frac12 I(A:R), \\
  R_B     &\geq \frac12 I(B:R), \\
  R_A+R_B &\geq \frac12 J(A:B) - E_{\text{sq}}(\rho_{AB}).
\end{align*}

While we are not able to describe the whole region 
for distributed compression, we provide the optimal 
sum of $R_A$ and $R_B$.  Our expression is not single-letter though. 

\begin{theorem}
\label{thm:main}
{\bf (Bipartite case)}
Given a bipartite source $\rho_{AB}$, the minimal sum of the rates of 
distributed compression is given by 
\be 
  R_A + R_B = \frac{1}{2}J(A:B) - \qp(\rho_{AB}).
  \label{eq:jqp}
\ee
\end{theorem}
\noindent
The direct and converse part of the theorem is proved in the Supplementary Materials.

\medskip\noindent
{\it Multipartite case.}
For $N$ parties $A_1\ldots A_N$, we define 
$J(A_1: \ldots :A_N) = S(A_1) + \ldots + S(A_N)+ S(A_1\ldots A_N)$.
We have the following formula  for the optimal sum of rates:
\be
  \sum_{i=1}^N R_i = \frac12 J(A_1:\ldots:A_N) - \qp(\rho).
\ee
The proof of the formula is analogous to the bipartite case. 
It is interesting to rewrite the formula 
in the exact case, i.e. if there is division 
$A_1\ldots A_N= \alpha_1 a_1 \ldots 
\alpha_N a_N$  such that the system $\alpha_1\ldots \alpha_N$ is product with $R$.
Then, 
\begin{equation}\begin{split}
  \sum_{i=1}^N R_i &= S(A_1\ldots A_N) \\
                   &\phantom{=}
                    + \frac12 \bigl[ I(A_1:\ldots : A_N)-I(\alpha_1:\ldots :\alpha_N) \bigr],
\end{split}\end{equation}
where we have divided the sum into two parts: 
$S(A_1\ldots A_N)$ is the rate when all the systems 
are together (this would be the rate of compression, if we were sending
classical information), while the second term is the quantum correction.
The latter says that in distributed case, we have to send all the correlations that are 
not independent of $R$: the total correlations 
$I(A_1:\ldots : A_N)$ minus  the independent ones 
$I(\alpha_1:\ldots :\alpha_N)$. 

\medskip\noindent
{\bf Conclusions.}
We have introduced a generalization of private bits, which we call
\qpriv, and defined the asymptotic amount of \qpriv\ as $\qp(\rho)$.
The quantity seems hard to compute, and even bounds are in general
hard to come by.

Apart from an upper bound by the squashed entanglement, we attempted
at giving another upper bound in terms of the relative entropy
of entanglement in the Supplementary Materials, which remains
contingent on an unproven Conjecture --
in fact, this conjecture itself is quite interesting as it claims
that the effect of entanglement locking~\cite{lock-ent} cannot occur if the
state is product.
Furthermore, we expressed our belief that \qpriv\ is generically zero
in another conjecture, about certain local algebras of operators
derived from the state.

The most important result is however the relation between
distributed quantum data compression and \qpriv: in fact, $\qp(\rho)$
is precisely by how much the optimal rate sum of the separate
compressors can go below $\frac{1}{2}J$, which is the rate
guaranteed by fully quantum state merging. Note however that
Theorem~\ref{thm:main} describes the rate region only in a very
weak sense: looking at the proof, we see of course that for a sequence
of isometric splittings $A^n \hookrightarrow a\alpha$, and $B^n \hookrightarrow b\beta$
with asymptotic \qpriv\ in $\alpha\beta$, one can achieve the rate pair
\begin{align*}
  R_A &= \lim_{n\rightarrow\infty} \frac{1}{2n} I(a:R^n B^n), \\
  R_B &= \lim_{n\rightarrow\infty} \frac{1}{2n} I(b:R^n \alpha|a).
\end{align*}
But what region these points span is completely unclear. For instance,
it is open whether the extreme ``corner'' points
\begin{align*}
  R_A &= \frac12 I(A:R),\ R_B = S(B)-\qp(\rho_{AB}), \text{ and} \\
  R_A &= S(A)-\qp(\rho_{AB}),\ R_B = \frac12 I(B:R),
\end{align*}
are achievable -- though this appears rather doubtful, from looking
at the proof of Theorem~\ref{thm:main}. (Note that by the state redistribution protocol, we also get information on how much entanglement between senders and receiver is required.)
Analogous results and similar open questions are also obtainable 
in the multipartite case. 
There are many other problems, which we haven't touched in this paper. For example, 
we have not analysed the scenario  where classical 
communication between parties is allowed, e.g. one way or two way.
In particular, it is intriguing whether allowing 
a sublinear amount of classical communication could eliminate 
the discontinuity of the quantity which most likely holds 
in the case with no communication. One may also examine the scenario, 
where Alice and Bob are allowed to share entanglement 
not only with the receiver, but also between themselves. We also do not know
the amount of mutual independence in the singular cases such as the maximally correlated state.
We have also not resolved
whether our technical assumption that
the size of the systems  $\alpha$ and $\beta$ scale
linearly with the system size is needed.

%The classical analogue tell us something, \textcolor{red}{but now %i forget what it is...}

%\textcolor{red}{{\tt What about free ebits between Alice and Bob?}}

\newpage

\section{Supplementary materials}

\section{Connection with pbits and characterization of states with \qpriv}

Pbits are quantum states $\rho_{ABA'B'}$ which exhibit one bit 
of perfect key after Alice and Bob perform 
local measurement on parts $AB$ of their systems. 
This means that the results are uncorrelated with $R$,
and moreover, they are maximally correlated.
Equivalently, one could define private states as such states
$\rho_{ABA'B'}$, for which after tracing out systems $A'B'$,
 the systems $AB$ already represent perfect key, i.e. measurement is not needed. 
 Clearly these can be obtained from the original pbits by applying {\it coherent} measurement. 
(See \cite{keyhuge,karol-PhD} for a discussion of equivalent forms 
of private states.)

In any QKD protocol, Alice and Bob end up with pbits. Moreover 
the distillable key (by means of two-way, one-way classical communication or even 
with no communication at all) is simply given by the rate of obtaining 
pbits from many copies of a given shared state.

Let us now suppose that we apply the measurement coherently,
i.e. apply c-not onto some ancillas $\alpha\beta$. 
Then the system $\alpha\beta$ is product 
with $R$, and has mutual information $1$. 
Note however, that this is often an underestimate.
E.g. the singlet state, which is a valid pbit, 
has 2 bits of \qpriv. The characterization of 
states containing \qpriv\ is in analogy to a similar characterization  of pbits
(the unitary below is analogous to the so-called {\it twisting}~\cite{keyhuge}) and is given by Proposition \prop3 of the main paper.  Namely:
\begin{propositionn}
\label{prop:uhlmann}
A state $\rho_{ABA'B'}$ has \qpriv\ in systems 
$AB$, if and only if there exists an isometry $U:{A'B'} \rightarrow {CD}$ 
such that  
\be
  (\1_{AB} \ot U_{A'B'}) \rho_{ABA'B'} (\1_{AB} \ot U_{A'B'})^\dagger
     = \psi_{ABC}\ot \rho_{D},
\ee
with a pure state $\psi$ on $ABC$.
\end{propositionn}
{\bf Proof.}
It evident that states satisfying the above conditions 
have \qpriv. Conversely, let us take a state 
which has \qpriv. We consider the
purification $\psi_{ABA'B'R}$  of $\rho_{ABA'B'}$ 
and by assumption know that $\rho_{ABR}=\rho_{AB}\ot \rho_R$,
where $\rho_{ABR}=\tr_{A'B'}\psi_{ABA'B'R}$.
Then one can clearly find a state of the form
$\psi_{ABC} \ot \varphi_{RD}$, which
is also a purification of $\rho_{ABR}$ (where $\varphi_{RD}$ 
is purification of $\rho_R$ and $\psi_{ABC}$ 
is purification of $\rho_{AB}$). Since purifications 
of a given state  are related by an isometry, we obtain that 
\be
  \psi_{ABC} \ot \varphi_{RD} 
      = (\1_{ABR} \ot U_{A'B'}) \psi_{ABA'B'R} (\1_{ABR} \ot U_{A'B'})^\dagger,
\ee
because both states are purifications of $\rho_{ABR}$,
as we wanted. \qed

The reasoning is thus similar to the decoupling technique \cite{SchumacherW01-approx}
which was succesfully applied to quantum state merging, 
both in the original \cite{SW-nature,sw-long} and 
fully quantum setting \cite{AbeyesingheDHW2006-fullSW} 
as well as for new proofs of 
quantum Shannon theorem, see e.g. \cite{HaydenHWY2008-Shannon}. 
The same technique 
was also used in \cite{keyhuge} in proving relations 
between two key distillation scenarios:
one related to distillation of pbits (i.e. states belonging to 
Alice and Bob solely) and the other one related to 
distillation of key as a tripartite state between Alice, Bob 
and Eve.

As a matter of fact this simple observation is what allows 
us to apply the notion of \qpriv\ to the problem of 
distributed compression of quantum information discussed further on in the paper.

\section{Discussion of Conjecture \con4 and discontinuity of mutual independence}

As a first step to proving Conjecture \con4 of the main paper, we can show that 
If $I_{ind}(\rho_{AB}) > 0$, then there exist operator subspaces ${\cal
A}^{(n)}$ and
 ${\cal B}^{(n)}$ of the full local operator algebras ${\cal L}(A^n)$
and ${\cal L}(B^n)$,
 respectively, both containing the unit $\1$, and a sequence of
 $\epsilon_n \rightarrow 0$, with the following property. For any ensemble
 decomposition of $\rho_{AB}^{\otimes n} = \sum_j p_j \proj{\psi_j^{(n)}}$
 and any $A\in{\cal A}^{(n)}$, $B\in{\cal B}^{(n)}$,
 \begin{equation}
  \sum_j p_j \bigl| \bra{\psi_j^{(n)}} A\otimes B \ket{\psi_j^{(n)}}
                         - \tr\rho^{\otimes n}(A\otimes B) \bigr|
  \leq \epsilon_n \| A \otimes B \|.
\label{eq:manyletter}
 \end{equation}
%%
%\begin{lemma}
%\label{lemma:manyletter}
%If $\rho_{AB}$ has $\qp(\rho_{AB}) > 0$
%then
%$\bra{\psi_j^{(n)}}A\otimes B \ket{\psi_j^{(n)}}-\bra{\psi_k^{(n)}}A\otimes B \ket{\psi_k^{(n)}}\leq \epsilon$  $\forall \ket{\psi_j^{(n)}},\ket{\psi_k^{(n)}}$ in the typical subspace 
%of  $\rho_{AB}^{\otimes n}$.  
%\end{lemma}
One then needs to show that this implies that
the single-letter version (Conjecture \con4), also holds.  
That Equation (\ref{eq:manyletter}) holds is almost trivial. Namely,
%and indeed, holds for operators of any form, %not just
%those of the form $A\otimes B$.
consider the CP map $\Lambda$ which generates
productness.  Independence of Eve requires that the final
state is the same regardless of initial state $\ket{\psi_j^{(n)}}$. Then  for any operator $A'\otimes B'$, the ability to generate productness implies
\beq
\tr (A'\otimes B'\Lambda(\proj{\psi_j^{(n)}}))
\approx
\tr (A'\otimes B'\Lambda(\proj{\psi_k}^{(n)})) 
\eeq
We consider the dual map to go to the Heisenberg picture
of operators $A'\otimes B'$.  This is still a product map i.e.
$\Lambda^*(A'\otimes B')=A\otimes B$, since $\Lambda$ was a product map.
So we have
\beq
\bra{\psi_j^{(n)}}A\otimes B \ket{\psi_j^{(n)}}
\approx
\bra{\psi_k^{(n)}}A\otimes B \ket{\psi_k^{(n)}}
\eeq
%
%Now, we can always choose $\ket{\psi_j^{(n)}}$ to be some tensor product state i.e. 
%some $\psi^1\otimes\psi^2...\psi^n$. 
%Then I think we just need to show that there must exist some $A'\otimes B'$ such that
%$\Lambda^*(A'\otimes B')$ is not trivial and single-letter (at least on one out of the $n$
%copies), 
%provided that $\Lambda$ is not trivial
%(i.e. complete tracing out).  Since we can write $\Lambda$ is 
%appending an ancilla (call it $\rho_{A'B'}$), 
%applying some
%$U_{AA'}\otimes U_{BB'}$, and tracing out, this amounts to showing that
%for all $U_{AA'}\otimes U_{BB'}$ we can find some operator $\alpha$ such that
%$U^\dagger_{AA'}\alpha \otimes \id_a U_{AA'}$ is not the identity when restricted
%to one of the copies.  Note that there may be some $\ket{\psi_k}$ which have vanishing probability
%in the ensemble, so should be careful...
%}

Observe that the conjecture holds for the
maximally correlated states of the example in the main paper, where $A\otimes B = \sigma_z\otimes\sigma_z$. 
Since, to get \qpriv\ from 
maximally correlated states Alice and Bob dephase their systems locally  in bases of $\sigma_z$ and apply privacy amplification to 
the resulting classical data, it suggests 
a general method for distilling \qpriv\ -- namely, both
parties measure or dephase a set of 
commuting $A\otimes \1$ and $\1\otimes B$, 
and apply local privacy amplification.
% as 
%in the example given in section \sec1a\ of the paper. 

I.e. suppose that Alice and Bob, 
by dephasing their state in local bases 
obtain a probability distribution, about which Eve has 
only local information i.e. Eve's information is a sole 
result of action of two local channels, one from Alice, a
second from Bob. 
Then it is clear that Alice and Bob simply need to 
perform local privacy amplification on their measurement 
outcomes.
Now, privacy amplification forces Alice to
loose $nI(A:E)$ systems, which will reduce her mutual information with Bob by the
same amount.  Likewise, Bob needs to loose $nI(B:E)$ systems, reducing the mutual
information between him and Alice by potentially an additional $nI(B:E)$ bits (although
potentially no reduction need occur, e.g. in the case of maximally correlated states).
This protocol will produce
mutual independence at a rate of at least
$I(\tilde{A}:\tilde{B})-I(\tilde{A}:E)-I(\tilde{B}:E)$, provided this quantity is positive. Here  $\tilde{A},\tilde{B}$
are the post measurement outcomes.  

To obtain an example, one could consider first the classical 
state  of the form 
\begin{equation}\begin{split}
\rho_{ABE} &= \sum_{i,j=1}^d p_{ij} |ij\>\<ij| \ot 
                \bigl( (1-p) \1_{E_1}/d + p |i\>\<i| \bigr) 
%           &\phantom{========}
                \ot \bigl( (1-p) \1_{E_2}/d + p |j\>\<j| \bigr).
\label{eq:local}
\end{split}\end{equation}
where $p_{ij}$ is an arbitrary probability distribution, and
the first system is on $AB$, and the second and third systems are
with Eve.
Here, obviously, Eve has only local information in the above sense.
Now, one can consider the purification, and hand it to 
Alice and Bob. It seems, that if we distribute it into 
Alice and Bob systems in a nontrivial way, the 
total state should not show \qpriv\ on the single copy level. 

However, one may find examples where this protocol may not work.
E.g. we start with
the example of equation \ref{eq:local} 
(purified in whichever way, with the purification handed 
to Alice and Bob).
%\beq
%\rho_{ABE}= \sum_{i,j=1}^d p_{ij} |ij\>\<ij| \ot 
%                \bigl( (1-p) \1_{E_1}/d + p |i\>\<i| \bigr)          
%                \ot \bigl( (1-p) \1_{E_2}/d + p |j\>\<j| \bigr).
%\nonumber
%\eeq
Then, as the operators $A$ and $B$ we can take operators of the form 
$X_A\ot \1_{A'}$ and $X_B\ot \1_{B'}$ where $X$'s are diagonal 
in the basis which is complementary to the standard basis. The results 
will then be completely uncorrelated. However, 
one can choose the operators to be diagonal in the standard 
basis, so that the method works. It is possible 
that we can always find such operators $A$ and $B$,
such that the measurement in their bases, and subsequent local 
privacy amplification is a good strategy (provided 
the mutual information between Alice and Bob 
is large enough). 

Finally, we wish to highlight that Conjecture \con4 refers to the impossibility
of obtaining a positive rate for \qpriv.  There are however situations where
one can obtain a single bit of \qpriv\ but at zero rate.  I.e. given an arbitrarily large number of copies of an initial system, one can get a single bit which has non-zero 
correlation and is private \cite{maj-withrenato}.

\section{Upper bound for \qpriv}
In~\cite[Theorem X.2]{AbeyesingheDHW2006-fullSW} one can find implicitly a proof
of the bound $\qp(\rho_{AB}) \leq E_{\text{sq}}(\rho_{AB})$,
the squashed entanglement~\cite{Winter-squashed-ent}.
The argument, in a nutshell, is this: isometrically splitting $A^n \rightarrow a\alpha$
and $B^n \rightarrow b\beta$, and for any likewise splitting
$R^n \rightarrow EF$, we assume approximate \qpriv\,
i.e.~$\rho_{\alpha\beta R^n} \approx \rho^{\alpha\beta} \ot \rho_{R^n}$,
which by Fannes' inequality~\cite{Fannes1973} (in the form given
in~\cite{Alicki-Fannes}) translates into
\be
  o(n) \geq I(\alpha\beta:R^n) \geq I(\alpha\beta:E),
\ee
the second inequality by monotonicity of the quantum mutual information.
(Note that in the inequality of~\cite{Alicki-Fannes} only a dimensional
factor of $\log|R^n| = O(n)$ enters.)
Hence, using a straightforward identity,
\begin{align}
  I(\alpha:\beta) &= I(\alpha:\beta|E) \nonumber\\
                  &\phantom{==}
                   + I(\alpha:E) + I(\beta:E) - I(\alpha\beta:E) \nonumber\\
                   \label{eq:product-cond-for-sq}
                  &\leq I(\alpha:\beta|E) + I(\alpha\beta:E)     \\
                  &\leq I(A^n:B^n|E) + o(n),
\end{align}
using monotonicity three more times. By the definition of squashed entanglement,
and its additivity~\cite{Winter-squashed-ent}, we find
\be
  \frac{1}{n}\frac{1}{2}I(\alpha:\beta) - o(1) \leq E_{\text{sq}}(\rho_{AB}),
\ee
and we only need to take the limit $n \rightarrow \infty$.  \qed

\medskip\noindent
{\bf Remark.} The above proof actually requires much less than 
$R^n$ to be (approximately) product with $\alpha\beta$: indeed,
looking at eq.~(\ref{eq:product-cond-for-sq}), we see that the correction
term could be replaced by any of $I(\alpha:E)$ or $I(\beta:E)$, so that
it is sufficient that either $\alpha$ or $\beta$ is product with $R^n$.
This suggests that the bound by squashed entanglement is not 
particularly tight.

\medskip

{\bf Proof.} 
Since mutual information is additive, it is enough to prove that $E_r$ is an upper bound. 
Then by applying the result to many identical copies it follows that $E_r^\infty$ is a bound, too. 
Consider the closest separable state $\sigma$ to $\rho$ in relative entropy distance,
with respect to the cut $AA':BB'$.
Since our state has \qpriv\ in system $AB$, according to Proposition \ref{prop:uhlmann}
there exists an isometry $U:{A'B'} \rightarrow {CD}$ such that 
\be
  (\1_{AB} \ot U_{A'B'})  \rho_{ABA'B'} (\1_{AB} \ot U_{A'B'})^\dagger = \psi_{ABC} \ot \rho_{D}.
\ee
Thus we have 
\begin{equation}\begin{split}
  E_r(\rho_{ABA'B'}) &=    S(\rho_{ABA'B'} \|\sigma_{ABA'B'})                    \\
                     &=    S(\psi_{ABC} \ot \rho_{D} \| \sigma_{ABCD}'),         \\
                     &\geq S(\psi_{ABC} \| \sigma_{ABC}')
  \label{eq:E_R:lowerbound}
\end{split}\end{equation}
where $\sigma'= (\1_{AB} \ot U_{A'B'}) \sigma (\1_{AB} \ot U_{A'B'})^\dagger$.
Note that while $\sigma_{AA'BB'}$ is separable, $\sigma'$ has
no obvious separability properties. However, since
$\sigma_{AA'BB'} = \sum_i p_i \xi^{(i)}_{AA'} \ot \eta^{(i)}_{BB'}$, we find
that
\(
  \sigma_{ABC}' = \sum_i p_i \sigma^{(i)}_{ABC},
\)
where each
\be
  \sigma^{(i)}_{ABC} = \tr_D  (\1_{AB} \ot U_{A'B'})
                                  \bigl(\xi^{(i)}_{AA'} \ot \eta^{(i)}_{BB'}\bigr)
                              (\1_{AB} \ot U_{A'B'})^\dagger
\label{eq:sigma_i}
\ee
has the property that the marginal $\sigma^{(i)}_{AB}$ is a product state.

Let us now imagine for a while that the system $C$ 
is distant from $A$.  We then take $k$ copies of the state $\psi$, 
and consider an operation that merges system $C$ to system $A$,
in such a way that the new state $\rho_{A^k C^k B^k}'$ satisfies 
\be
  \bigl\| \rho_{A^k C^k B^k}' - \psi_{ACB}^{\ot k} \bigr\|_1 \leq \ep.
  \label{eq:close}
\ee
It is known, that one can perform 
such an operation~\cite{AbeyesingheDHW2006-fullSW} by sending 
$\frac{1}{2} I(C:B) + \delta$  qubits per copy, with $\ep$ and 
$\delta$ tending to zero for large $k$. 

Before proceeding, we shall quickly outline the further reasoning. 
As a result of merging (i.e. enlarging the cut $A:B$ to $AC:B$), we
arrive at a pure state with entanglement $S(B)$ per copy.
We shall apply the same operation to $\sigma'$, and appealing to
Conjecture \conjnolock for the state $\sigma^{(i)}$, will argue
that it can increase its entanglement by no more than the number of sent qubits.
Thus we are left with relative entropy between a pure state 
of given entanglement and some state which is not separable anymore,  
but entanglement of which is bounded from above by $\frac{1}{2} I(C:B)$ per copy. 
Therefore, the relative entropy must be bounded from below, and 
we shall show that it is bounded by the difference between the 
entanglements of the two states, which is precisely $\frac{1}{2} I(A:B)$.
%\textcolor{red}{
Coming back to the proof, let us examine the action of the merging protocol on 
the state $(\sigma_{ACB}')^{\ot k}$. The latter is a mixture of 
states $\sigma^{(i_1)}_{ACB}\ot \ldots \ot \sigma^{(i_k)}_{ACB}$ 
as in (\ref{eq:sigma_i}), which are product after tracing out
the systems $C$. 
Therefore, by Conjecture \conjnolock\,
after merging, for each such state the logarithmic negativity $E_N$ 
cannot become greater than  $k(I(C:B)+\delta)$ or,  equivalently, 
$\| (\cdot)^\Gamma \|_1 \leq 2^{k(I(C:B)+\delta)}$ 
Due to convexity of the trace norm, we obtain that the same 
is true for the total state, hence, by monotonicity of logarithm,
$E_N\leq k(I(C:B)+\delta)$ for the total state.
%}
%the logarythmic negativity \ref{logneg}
%is bounded by $k(I(A':B)+\delta)$ for the mixture of such states. 
%It is because, as one can show, $E_N\leq \log \esr$. We note here 
%that we couldn't use directly logarythmic negativity
%instead of $\esr$, because the former is not convex.

Continuing from eq.~(\ref{eq:E_R:lowerbound}), and denoting $\tilde{A} = AC$, 
we obtain
\be
  k E_r(\rho_{ABA'B'}) \geq S(\rho_{\tilde{A}^k B^k} \| \sigma_{\tilde{A}^k B^k}''),
  \label{eq:after-merging}  
\ee
where $\rho_{\tilde{A}^k B^k}'$ satisfies eq.~(\ref{eq:close}), while,
as argued above, for the state $\sigma_{\tilde{A}^k B^k}''$ 
resulting from the merging applied to
$(\sigma_{ACB}')^{\ot k}$, we have
\be
  E_N(\sigma'') \leq k \bigl( I(C:B)+\delta \bigr). 
  \label{eq:en}
\ee
We now apply entanglement concentration \cite{BBPS1996} to the state 
$\rho_{\tilde{A}^k B^k}'$, which turns it to a state
close to the maximally entangled state living on 
dimension $d\times d $ with $d\geq 2^{k\bigl(S(B)-\delta'\bigr)}$.
The same operation is applied to state $\sigma_{\tilde{A}^k B^k}''$. 
This operation we follow by $U \ot U^*$ twirling~\cite{Werner1989,reduction}. 
As a result, we obtain two isotropic states $\rho_{\text{iso}}(F,d)$, $\sigma_{\text{iso}}(F',d)$ 
where we use the notation
\be
  \rho_{\text{iso}}(F,d)=F \Phi_d + (1-F) \frac{(\1-\Phi_d)}{d^2-1},
\ee
with the maximally entangled state $\Phi^d$
on a $d\times d$-system. The dimension $d$ satisfies $d\geq 2^{k (S(B) - \delta')}$ 
where we can take $\delta' \to 0$ for large $k$. Moreover $F\to 1$ for large $k$, 
and $E_N(\sigma_{\text{iso}})\leq k(I(A':B)+\delta)$, again with $\delta\to 0$ for large $k$. 

Since for $F'\geq \frac 1d$, we have $E_N\bigl(\sigma_{\text{iso}}(F',d)\bigr)=\log (F'd)$,
which by eq.~(\ref{eq:en}) gives 
\be
  \log F' \leq k \left(\frac12 I(C:B) - S(B) + \delta +\delta' \right).
\ee
Since the state $\psi_{ACB}$ was pure, we obtain
\be
  \log F' \leq - k \left(\frac12 I(A:B) - \delta - \delta' \right).
\ee
The monotonicity of relative entropy and eq.~(\ref{eq:after-merging}) now give
\begin{equation}\begin{split}
  k E_r(\rho_{AA'BB'}) &\geq S\bigl(\rho_{\text{iso}}(F,d)\|\sigma_{\text{iso}}(F',d)\bigr) \\
                       &=    S\bigl(\{F,1-F\}\|\{F',1-F'\}\bigr).
\end{split}\end{equation}
This expression is continuous in $F$ for $F'<1$, hence we can set $F=1$, incurring
another small deviation, 
$S\bigl(\{F,1-F\}\|\{F',1-F'\}\bigr) \geq -\log F'-\delta''$,
and we obtain
\be
  E_r(\rho_{AA'BB'})\geq \frac12 I(A:B) - \delta - \delta' -\delta''.
\ee 
Sending $\delta$, $\delta'$ and $\delta''$ to zero, we obtain the result.  \qed

\section{Proof of Direct Part and Converse of Theorem 5}
\label{sec:proofs}
\noindent
{\bf Remark.} The converse part for the case without 
shared entanglement was provided implicitly in~\cite{AbeyesingheDHW2006-fullSW}.
Here we show that it extends to our scenario of auxiliary ebits.

\medskip\noindent
%\medskip\noindent
{\bf Proof (Direct part).} 
Let us first note that to achieve the goal it is enough that at the end 
Charlie's system and the system R are in a pure state. Indeed, all purifications of $R$ 
are equivalent up to an isometry on Charlie side. Applying such an isometry, 
Charlie can reconstruct the required state.

To proceed, let us first assume
that the state $\rho_{AB}^{\ot m}$ has exact \qpriv. That is, 
there are local isometries which allow us to write the state as 
$\rho_{a\alpha b \beta}$ with $\alpha \beta$ being private,
i.e. $I(\alpha\beta: R^m)=0$.  

Let us now argue that it is enough 
to send $a$ and $b$ to Charlie. Indeed, suppose that Charlie 
has the $ab$ system. Then, due to Proposition \ref{prop:uhlmann},
Charlie can apply an isometry $U:ab \rightarrow C\gamma$ 
such that $\alpha\beta\gamma$ is in a pure state. 
Now he can remove system $\gamma$ because
it is not correlated with $R$, and the state of the system 
$R^mC$ is now pure. Thus, given our initial remark, the transmission 
of information to Charlie was achieved. 

Now, let us argue that we can achieve the rate sum in eq.~(\eqjqp) of the main paper.
\be 
  R_A + R_B = \frac{1}{2}J(A:B) - \qp(\rho_{AB}).
 \nonumber
\ee
To this end, Alice and Bob will send $a$ and $b$ by way of the
state redistribution protocol~\cite{DevetakY2006-cond-info}. 
Suppose that Alice sends $a$, using $\alpha$ as
side-information to be retained at her side; then Bob can send
$b$ using $\beta$ as side-information to be retained by him,
and $a$ as side-information at the receiver
(the other case gives the same sum of rates). Thus, we obtain
the rate pair
\begin{align}
  \label{eq:RA}
  mR_A &= \frac12 I(a:R^mb\beta)          \nonumber \\
       &= \frac12 \bigl( S(a)+S(A^m)-S(\alpha) \bigr), \\
  \label{eq:RB}
  mR_B &=\frac12 I(b:R^m\alpha|a)         \nonumber \\                                  
       &=\frac12 \bigl( S(ab)+S(R^mA^m)-S(\beta)-S(a) \bigr),
\end{align}
so that 
\be
  m(R_A+R_B) = \frac12 \bigl( S(A^m)+S(B^m)+S(ab)-S(\alpha)-S(\beta) \bigr).
  \label{eq:sum}
\ee
Now, since the system $\alpha\beta$ is product with $R^m$, we have 
$I(\alpha\beta:R^m)=0$, i.e.~$S(A^mB^m)+S(\alpha\beta)=S(ab)$.
Inserting this into eq.~(\ref{eq:sum}) and dividing by $m$,
we obtain eq.~(\eqjqp) of the main paper. 

Suppose now that \qpriv\ is not exact, but achievable asymptotically.
This means that given any $\epsilon$, 
there some number of copies $m$ such that Alice and Bob can transform the state 
$\psi_{ABR}^{\ot m}$ via local isometries into
a state $\psi_{a\alpha b\beta R}^{(m)}$ whose reductions satisfy
\be
  \bigl\| \rho^{(m)}_{\alpha \beta R}-\rho^{(m)}_{\alpha \beta}\ot\rho_R^{\ot m} \bigr\|_1 \leq \ep.
  \label{eq:approx-indep}
\ee

If $ab$ could be sent to Charlie exactly, then he would  
be able to reconstruct the purification of $R$ 
with fidelity $\geq \text{const.}\sqrt{\ep}$. 
To do this would be too costly in terms of communication, 
however they can be transferred with arbitrary 
high fidelity, by use of state redistribution,
applied to $k$ copies of $\rho^{(m)}_{a\alpha b\beta}$,
where $k$ is chosen to be large, but say $k\leq 1/\sqrt{\ep}$, so that
in light of eq.~(\ref{eq:approx-indep}),
\be
  \bigl\| \rho^{(m)\ot k}_{\alpha \beta R}
              - \rho^{(m)\ot k}_{\alpha \beta}\ot\rho_R^{\ot km} \bigr\|_1 \leq \sqrt{\ep}.
  \label{eq:approx-indep-k}
\ee
Then, having $ab$ with high fidelity,
Charlie can recover the source with arbitrarily high fidelity.
Letting $m \rightarrow\infty$ and $\ep \rightarrow 0$ would complete the
direct part. All that is left is to argue why a comparably small $k$
can suffice -- note that theorems such as the main result 
in~\cite{DevetakY2006-cond-info} are typically stated as asymptotic
results for $k \rightarrow \infty$ as the source state remains
fixed.

We use the insight of~\cite{Jono-coherent-relay} to see that
only two one-shot versions of the coherent state merging
protocol~\cite{AbeyesingheDHW2006-fullSW} are needed. The
fidelity of those, in turn, only depends on the fidelity, dimensions
and maximum eigenvalue of certain typical subspaces. If we use
the \emph{entropy-typical subspaces} of~\cite{Schumacher1995},
i.e.~for $\sigma = \sum_{x\in{\cal X}} \lambda_x |x\rangle\langle x|$
it is
\be
  {\cal S} = \text{span}\left\{ |x_1\ldots x_k\rangle : 
                     \left| \sum_{j=1}^k -\log\lambda_{x_j} - k S(\sigma) \right| \leq k\delta \right\},
\ee
and the projector $\Pi$ satisfies
\begin{align}
  \tr \sigma^{\ot k}\Pi           &\geq 1-\frac{2(\log|{\cal X}|)^2}{k\delta^2} =: 1-\eta, \\
  (1-\eta) 2^{kS(\sigma)-k\delta} &\leq \tr\Pi \leq 2^{kS(\sigma)+k\delta}, \\
  \Pi\sigma^{\ot k}\Pi            &\leq 2^{-kS(\sigma)-k\delta}.
\end{align}
We now apply this to $\rho^{(m)}_{ab\alpha\beta}$ and with $\delta = m\delta_0$.
So, mindful of eq.~(\eqqpriv-dimensionbound) of the main paper, we find
$\eta = \frac{O\left((\log c)^2\right)}{k\delta_0^2}$ above, while the
exponential rates -- normalized with $n=km$ -- are bounded within
$\delta_0$ around the entropy rate $\frac{1}{m} S(\sigma)$.
Choosing $\delta_0$ arbitrarily small, as a function of $k$,
say $\delta_0 = k^{-1/3}$, concludes the direct part.

\noindent
{\bf (Converse part).}
We shall actually study the whole rate region.
This is very similar to~\cite{AbeyesingheDHW2006-fullSW}, only now
we have to deal with the free entanglement; on the other hand, the 
converse also becomes easier since we are not after a single-letter
formulation.

We consider the most general protocol: for $n$ copies of the source,
initially $\psi_{ABR}^{\ot n}$ is distributed between Alice, Bob and
the reference. In addition, Alice and Charlie share entanglement
$\phi_{A_0C_a}$, Bob and Charlie share entanglement $\theta_{B_0C_b}$,
so that the state at the beginning is
\be
  \Psi_{A^nA_0,B^nB_0,R^n,C_a C_b} = \psi_{ABR}^{\ot n} \ot \phi_{A_0C_a} \ot \theta_{B_0C_b}.
\ee
Alice's (Bob's) encoding can be represented in the Stinespring form
as an isometry $A^n A_0 \rightarrow a_0\alpha$ ($B^n B_0 \rightarrow b_0\beta$);
Alice (Bob) then sends $a_0$ ($b_0$) to Charlie, keeping $\alpha$ ($\beta$).

Now the existence of a decoding operation of Charlie's, i.e.~a isometry
$a_0 b_0 C_a C_b \rightarrow A^n B^n \gamma$, such that the resulting
state $\widetilde\rho_{A^n B^n R^n} \approx \psi_{ABR}^{\ot n}$, implies -- indeed
is equivalent to -- 
$\rho^{(n)}_{\alpha\beta R^n} \approx \rho^{(n)}_{\alpha\beta} \ot \rho_R^{\ot n}$.
I.e.~there is approximate \qpriv\ in $\rho_{AB}^{\ot n}$.
The entanglement with Charlie in the procedure is of no consequence
here. Indeed, letting $a := a_0 C_a$ and $b := b_0 C_b$, we get
the following lower bounds on the rates:
\begin{align*}
  nR_A &\geq S(a_0) \geq \frac{1}{2} I(a_0:R^n B^n|C_a) 
        =                \frac{1}{2} I(a:R^n B^n),  \\
  nR_B &\geq S(b_0) \geq \frac{1}{2} I(b_0:R^n \alpha|a C_b)                                         
        =                \frac{1}{2} I(b:R^n \alpha|a),
\end{align*}
as $I(C_a:R^nB^n) = I(C_b:R^nA^n) = 0$.
So, as before in the direct part, we obtain the rate sum
\be\begin{split}
  n(R_A+R_B) &\geq \frac{1}{2}\bigl[ S(A^n)+S(B^n) \bigr.\\
             &\phantom{====}
                              \bigl. +S(ab)-S(\alpha)-S(\beta) \bigr] \\
             &\geq \frac{1}{2}J(A^n:B^n) - \frac{1}{2}I(\alpha:\beta) - o(n),
\end{split}\ee
using the fact that $R^n$ and $\alpha\beta$ are almost product,
i.e., invoking Fannes' inequality~\cite{Fannes1973} in the formulation
of~\cite{Alicki-Fannes}, $I(R^n:\alpha\beta) = o(n)$.
Taking $n\rightarrow\infty$ concludes the proof. \qed 

\medskip
Our theorem implies that one can beat the rate 
$\frac12 J(A:B) - D_0(\rho_{AB})$ of~\cite{AbeyesingheDHW2006-fullSW} where $D_0$ 
is the distillable entanglement by means of local operations only,
as had been suggested in~\cite{AbeyesingheDHW2006-fullSW}. 
Indeed, for pdits $\gamma$,
\be
  \qp(\gamma) \geq \frac{1}{2}\log d,
\ee
while there are pdits such that even $D^\leftrightarrow$ ($\geq D_0$)
is arbitrarily close to zero~\cite{pptkey}.

%\medskip\noindent
\section{Classical analogue}
\label{sec:classical-analogue}
In the paper, we have considered the case of the 
quantum mutual independence, as well as its relation 
with distributed compression.  We can also consider the analogous
classical problem.  Here, we will find that for classical
distributed compression, the solution is singular, as we suspect it is in the quantum case. 
 However, this does not imply that
mutual independence is singular, because as we shall see, in the classical case,
distributed compression and mutual independence are not as closely linked as 
they are in the quantum case.

For classical mutual independence, instead of a 
tripartite pure state, one considers a 
tripartite classical probability distribution $P_{XYZ}$ of random variables
$XYZ$, with $Z$ being
the reference and $XY$ being held by Alice and Bob. The probability that
the source produces $XYZ=xyz$ is denoted by $P_{XYZ}(xyz)$.
The definitions of
quantum mutual independence and distributed compression then follow 
exactly as in the quantum case. Note that the objective in the latter is
to allow Charlie to recreate a sample from the joint distribution
$P_{XYZ}$ (while $Z$ remains hidden and with the reference at all times),
not necessarily to reproduce the sample given initially to Alice and Bob.  
For distributed compression, the situation is then
analogous to blind compression with mixed states.  I.e. conditioned
on each $Z=z$, Alice and Bob are given a sample from the distribution $P_{XY|Z=z}$.
The extreme case that $Z=XY$ is a normal data compression problem,
since $P_{XY|Z=xy} = \delta_{XY,xy}$; it was solved by 
Slepian and Wolf~\cite{slepian-wolf} who showed that the rate sum $R_A+R_B$
can achieve the Shannon entropy $H(XY) = -\sum_{xy}P_{XY}(xy)\log P_{XY}(xy)$.

Of course, if $Z$ does not represent full information of $XY$, the optimal rate
sum could be smaller. We find below that for distributed
compression, the optimal rate sum is singular in the sense that 
for it to be smaller than $H(XY)$, the single-copy distribution
$P_{XYZ}$ must be in a certain set of measure zero. In particular,
the minimum rate sum is discontinuous.

To show this, we now ask what is the 
best compression rate which can be achieved in the case
of signal states which are mixed, even when Alice and Bob are together.  
Consider the distribution given to the senders for each $Z$; then there
exists a natural decomposition into a part which can depend on $Z$, 
and one which does not (i.e. a part which is redundant): there exists
a 1-1 identification
$\tau: {\cal X}\times{\cal Y} \rightarrow \bigcup_{\ell}^{.} {\cal J}_\ell \times {\cal K}_\ell$
such that for $\tau(xy) = jk \in {\cal J}_\ell \times {\cal K}_\ell$,
\beq
  P_{XY|Z=z}(xy) = q(\ell|z) P_{J|Z=z,L=\ell}(j) P_{K|L=\ell}(k),
  \label{eq:redundant-decomposition}
\eeq
with a distribution $q(\ell|z)$ for every $z$.
Then the best achievable compression rate is given by $H(LJ)$. I.e.,
the best one can do is to remove the {\it redundant part} $PK$ that
manifest already on the single-copy level -- collective actions
cannot do better than this.  The proof of this statement follows from a
straightforward application of the analogous quantum result of Koashi and
Imoto~\cite{KoashiI-compr} and we will therefore not give it here.

Now, when Alice and Bob are seperated and are attempting to perform
the compression, they may not even be able to achieve the rate
$R_A+R_B=H(LJ)$, because they may not be able to remove the redundant
part when they are in distant labs.  However, it still does provide a lower
bound on their rate, showing that only if there is single-copy redundancy in
$P_{XYZ}$, they can ever beat the rate sum $H(XY)$.
This leaves open the interesting problem of what
the achievable rate is -- we conjecture that it is possible to
go below $H(XY)$ if and only
if a decomposition into relevant and redundant parts as in
eq.~(\ref{eq:redundant-decomposition}) is achievable on the
single-copy level by local actions of Alice and Bob.

Note however, that in the classical case, the relevant quantity is how
much of the redundancy in the distribution -- that which is independent
of $Z$ -- can be removed.  The redundant part  is independent of $Z$
but the converse need not be true -- a random variable can be independent of $Z$
but not be redundant.
An example of this is the following distribution:
with some probability $p$, Alice and Bob have correlated bits (c), 
and with probabilty $1-p$, they have anti-correlated bits (a).  
I.e. ~$P_{XYZ}(00c)=P_{XYZ}(11c)=p/2$ and $P_{XYZ}(01a)=P_{XYZ}(10a)=(1-p)/2$.  
Here, both $X$ and $Y$ are independent of $Z$ but we conjecture that neither is redundant (in single copy it is obvious, 
but for collective actions it is not proven yet).  On the other 
hand, an operation on both $XY$ and remove a redundant bit, under the map $00\rightarrow 0$,
$11\rightarrow 0$, $01\rightarrow 1$ and $10\rightarrow 1$.

By contrast, \qpriv\ is defined as in the quantum case, only without
the factor of $1/2$: we consider local randomized functions $F$ and $G$
such that $F(X^n)G(Y^n)$ are
jointly asymptotically independent of $Z^n$, and
maximize the limiting rate 
\be
  \liminf_{n\rightarrow\infty} \frac{1}{n}I\bigl(F(X^n):G(Y^n)\bigr)
\ee
over all protocols, to obtain $\qp(X:Y|Z)$.
Observe that there are distributions with positive
\qpriv, but no gain in distributed compression. For example, 
let $X=Y$ and the conditional distribution of $X|Z$ be such that
it doesn't have any redundancy in the sense of 
eq.~(\ref{eq:redundant-decomposition}) and~\cite{KoashiI-compr}.
Then -- via local hashing by the same function --,
$\qp(X:Y|Z) = H(X|Z)$, but $R_A+R_B \geq H(X)$.

%\textcolor{red}{{\tt Do we have a proof that collective actions 
%cannot remove generic single letter redundany?}}

Let us finally remark, that one can 
unify quantum and classical approches as in \cite{EkertCHHOR2006-ABEkey} by considering mixed tripartite 
state $\rho_{ABE}$, and extend the definition in most natural way.

%\textcolor{green}{%
%\begin{itemize}
%  \item {\tt Dimension bound on $ab\alpha\beta$?}
%  \item {\tt Page 4: convex?}
%  \item {\tt Classical analogue and Conclusions!?}
%  \item {\tt Evidence for the Conjectures?}
%\end{itemize}}

\medskip\noindent 
{\bf Acknowledgments.}
We thank Daniel Gottesman, Aram Harrow and Avinatan Hassidim
for interesting discussions. We also thank Marco Piani 
for providing counterexamples to an earlier version of 
Conjecture 1 in the Supplementary Materials,
and Guillaume Aubrun for related communications.
MH and JO are supported by EC IP SCALA. 
JO and AW were supported by the Royal Society, and by EU grant QAP.
AW was also supported by U.K. EPSRC and
a Philip Leverhulme Prize. 
The Centre for
Quantum Technologies is funded by the Singapore Ministry of Education
and the National Research Foundation as part of the Research Centres
of Excellence programme. 
Part of the work was done at the University of Cambridge, 
the National Quantum Information Centre 
of Gda\'nsk and the Primrose Cafe, Bristol.

\bibliographystyle{apsrev}

%\bibliography{rmp12-hugekey,../refjono2}
%\bibliography{rmp13-hugekey}

\begin{thebibliography}{31}
\expandafter\ifx\csname natexlab\endcsname\relax\def\natexlab#1{#1}\fi
\expandafter\ifx\csname bibnamefont\endcsname\relax
  \def\bibnamefont#1{#1}\fi
\expandafter\ifx\csname bibfnamefont\endcsname\relax
  \def\bibfnamefont#1{#1}\fi
\expandafter\ifx\csname citenamefont\endcsname\relax
  \def\citenamefont#1{#1}\fi
\expandafter\ifx\csname url\endcsname\relax
  \def\url#1{\texttt{#1}}\fi
\expandafter\ifx\csname urlprefix\endcsname\relax\def\urlprefix{URL }\fi
\providecommand{\bibinfo}[2]{#2}
\providecommand{\eprint}[2][]{\url{#2}}

\bibitem[{\citenamefont{Bennett and Brassard}(1984)}]{BB84}
\bibinfo{author}{\bibfnamefont{C.~H.} \bibnamefont{Bennett}} \bibnamefont{and}
  \bibinfo{author}{\bibfnamefont{G.}~\bibnamefont{Brassard}}, in
  \emph{\bibinfo{booktitle}{Proceedings of the IEEE International Conference on
  Computers, Systems and Signal Processing}} (\bibinfo{publisher}{IEEE Computer
  Society Press, New York}, \bibinfo{address}{Bangalore, India, December 1984},
  \bibinfo{year}{1984}), pp. \bibinfo{pages}{175--179}.

\bibitem[{\citenamefont{Ekert}(1991)}]{Ekert91}
\bibinfo{author}{\bibfnamefont{A.~K.} \bibnamefont{Ekert}},
  \bibinfo{journal}{Phys. Rev. Lett.} \textbf{\bibinfo{volume}{67}},
  \bibinfo{pages}{661} (\bibinfo{year}{1991}).

\bibitem[{\citenamefont{Horodecki
  et~al.}(2005{\natexlab{a}})\citenamefont{Horodecki, Horodecki, Horodecki, and
  Oppenheim}}]{pptkey}
\bibinfo{author}{\bibfnamefont{K.}~\bibnamefont{Horodecki}},
  \bibinfo{author}{\bibfnamefont{M.}~\bibnamefont{Horodecki}},
  \bibinfo{author}{\bibfnamefont{P.}~\bibnamefont{Horodecki}},
  \bibnamefont{and}
  \bibinfo{author}{\bibfnamefont{J.}~\bibnamefont{Oppenheim}},
  \bibinfo{journal}{Phys. Rev. Lett.} \textbf{\bibinfo{volume}{94}},
  \bibinfo{pages}{160502} (\bibinfo{year}{2005}{\natexlab{a}}),
  \eprint{quant-ph/0309110}.

\bibitem[{\citenamefont{Horodecki
  et~al.}(2005{\natexlab{b}})\citenamefont{Horodecki, Horodecki, Horodecki, and
  Oppenheim}}]{keyhuge}
\bibinfo{author}{\bibfnamefont{K.}~\bibnamefont{Horodecki}},
  \bibinfo{author}{\bibfnamefont{M.}~\bibnamefont{Horodecki}},
  \bibinfo{author}{\bibfnamefont{P.}~\bibnamefont{Horodecki}},
  \bibnamefont{and} \bibinfo{author}{\bibfnamefont{J.}~\bibnamefont{Oppenheim}}
  (\bibinfo{year}{2005}{\natexlab{b}}), \eprint{quant-ph/0506189}.

\bibitem[{\citenamefont{Buscemi}(2009)}]{buscemi-shredding}
\bibinfo{author}{\bibfnamefont{F.}~\bibnamefont{Buscemi}}
  (\bibinfo{year}{2009}), \eprint{arXiv:0807.3594}.

\bibitem[{\citenamefont{Abeyesinghe et~al.}(2006)\citenamefont{Abeyesinghe,
  Devetak, Hayden, and Winter}}]{AbeyesingheDHW2006-fullSW}
\bibinfo{author}{\bibfnamefont{A.}~\bibnamefont{Abeyesinghe}},
  \bibinfo{author}{\bibfnamefont{I.}~\bibnamefont{Devetak}},
  \bibinfo{author}{\bibfnamefont{P.}~\bibnamefont{Hayden}}, \bibnamefont{and}
  \bibinfo{author}{\bibfnamefont{A.}~\bibnamefont{Winter}}
  (\bibinfo{year}{2006}), \eprint{quant-ph/0606225}.

\bibitem[{\citenamefont{Christandl and Winter}(2004)}]{Winter-squashed-ent}
\bibinfo{author}{\bibfnamefont{M.}~\bibnamefont{Christandl}} \bibnamefont{and}
  \bibinfo{author}{\bibfnamefont{A.}~\bibnamefont{Winter}},
  \bibinfo{journal}{J. Math. Phys.} \textbf{\bibinfo{volume}{45}},
  \bibinfo{pages}{829} (\bibinfo{year}{2004}), \eprint{quant-ph/0308088}.

\bibitem[{\citenamefont{Vedral et~al.}(1997)\citenamefont{Vedral, Plenio,
  Rippin, and Knight}}]{VPRK1997}
\bibinfo{author}{\bibfnamefont{V.}~\bibnamefont{Vedral}},
  \bibinfo{author}{\bibfnamefont{M.~B.} \bibnamefont{Plenio}},
  \bibinfo{author}{\bibfnamefont{M.~A.} \bibnamefont{Rippin}},
  \bibnamefont{and} \bibinfo{author}{\bibfnamefont{P.~L.}
  \bibnamefont{Knight}}, \bibinfo{journal}{Phys. Rev. Lett.}
  \textbf{\bibinfo{volume}{78}}, \bibinfo{pages}{2275} (\bibinfo{year}{1997}),
  \eprint{quant-ph/9702027}.

\bibitem[{\citenamefont{Vedral and Plenio}(1998)}]{PlenioVedral1998}
\bibinfo{author}{\bibfnamefont{V.}~\bibnamefont{Vedral}} \bibnamefont{and}
  \bibinfo{author}{\bibfnamefont{M.~B.} \bibnamefont{Plenio}},
  \bibinfo{journal}{Phys. Rev. A} \textbf{\bibinfo{volume}{57}},
  \bibinfo{pages}{1619} (\bibinfo{year}{1998}), \eprint{quant-ph/9707035}.

\bibitem[{\citenamefont{Christandl and Winter}(2005)}]{ChristandlW-lock}
\bibinfo{author}{\bibfnamefont{M.}~\bibnamefont{Christandl}} \bibnamefont{and}
  \bibinfo{author}{\bibfnamefont{A.}~\bibnamefont{Winter}},
  \bibinfo{journal}{IEEE Trans. Inf. Theory} \textbf{\bibinfo{volume}{51}},
  \bibinfo{pages}{3159} (\bibinfo{year}{2005}), \eprint{quant-ph/0501090}.

\bibitem[{\citenamefont{\.Zyczkowski et~al.}(1998)\citenamefont{\.Zyczkowski,
  Horodecki, Sanpera, and Lewenstein}}]{ZyczkowskiHSP-vol}
\bibinfo{author}{\bibfnamefont{K.}~\bibnamefont{\.Zyczkowski}},
  \bibinfo{author}{\bibfnamefont{P.}~\bibnamefont{Horodecki}},
  \bibinfo{author}{\bibfnamefont{A.}~\bibnamefont{Sanpera}}, \bibnamefont{and}
  \bibinfo{author}{\bibfnamefont{M.}~\bibnamefont{Lewenstein}},
  \bibinfo{journal}{Phys. Rev. A} \textbf{\bibinfo{volume}{58}},
  \bibinfo{pages}{883} (\bibinfo{year}{1998}), \eprint{quant-ph/9804024}.

\bibitem[{\citenamefont{Vidal and Werner}(2002)}]{Vidal-Werner}
\bibinfo{author}{\bibfnamefont{G.}~\bibnamefont{Vidal}} \bibnamefont{and}
  \bibinfo{author}{\bibfnamefont{R.~F.} \bibnamefont{Werner}},
  \bibinfo{journal}{Phys. Rev. A} \textbf{\bibinfo{volume}{65}},
  \bibinfo{pages}{032314} (\bibinfo{year}{2002}), \eprint{quant-ph/0102117}.

\bibitem[{\citenamefont{Horodecki
  et~al.}(2005{\natexlab{c}})\citenamefont{Horodecki, Horodecki, Horodecki, and
  Oppenheim}}]{lock-ent}
\bibinfo{author}{\bibfnamefont{K.}~\bibnamefont{Horodecki}},
  \bibinfo{author}{\bibfnamefont{M.}~\bibnamefont{Horodecki}},
  \bibinfo{author}{\bibfnamefont{P.}~\bibnamefont{Horodecki}},
  \bibnamefont{and}
  \bibinfo{author}{\bibfnamefont{J.}~\bibnamefont{Oppenheim}},
  \bibinfo{journal}{Phys. Rev. Lett.} \textbf{\bibinfo{volume}{94}},
  \bibinfo{pages}{200501} (\bibinfo{year}{2005}{\natexlab{c}}),
  \eprint{quant-ph/0404096}.

\bibitem[{\citenamefont{Horodecki}(2001)}]{Michal2001}
\bibinfo{author}{\bibfnamefont{M.}~\bibnamefont{Horodecki}},
  \bibinfo{journal}{Quantum Inf. Comp.} \textbf{\bibinfo{volume}{1}},
  \bibinfo{pages}{3} (\bibinfo{year}{2001}).

\bibitem[{\citenamefont{Horodecki}(2008)}]{karol-PhD}
\bibinfo{author}{\bibfnamefont{K.}~\bibnamefont{Horodecki}}
  (\bibinfo{year}{2008}).

\bibitem[{\citenamefont{Schumacher and
  Westmoreland}(2001)}]{SchumacherW01-approx}
\bibinfo{author}{\bibfnamefont{B.}~\bibnamefont{Schumacher}} \bibnamefont{and}
  \bibinfo{author}{\bibfnamefont{M.~D.} \bibnamefont{Westmoreland}},
  \bibinfo{journal}{Quantum Information Processing}
  \textbf{\bibinfo{volume}{1}}, \bibinfo{pages}{5} (\bibinfo{year}{2001}),
  \eprint{quant-ph/0112106}.

\bibitem[{\citenamefont{Horodecki
  et~al.}(2005{\natexlab{d}})\citenamefont{Horodecki, Oppenheim, and
  Winter}}]{SW-nature}
\bibinfo{author}{\bibfnamefont{M.}~\bibnamefont{Horodecki}},
  \bibinfo{author}{\bibfnamefont{J.}~\bibnamefont{Oppenheim}},
  \bibnamefont{and} \bibinfo{author}{\bibfnamefont{A.}~\bibnamefont{Winter}},
  \bibinfo{journal}{Nature} \textbf{\bibinfo{volume}{436}},
  \bibinfo{pages}{673} (\bibinfo{year}{2005}{\natexlab{d}}),
  \eprint{quant-ph/0505062}.

\bibitem[{\citenamefont{Horodecki et~al.}(2007)\citenamefont{Horodecki,
  Oppenheim, and Winter}}]{sw-long}
\bibinfo{author}{\bibfnamefont{M.}~\bibnamefont{Horodecki}},
  \bibinfo{author}{\bibfnamefont{J.}~\bibnamefont{Oppenheim}},
  \bibnamefont{and} \bibinfo{author}{\bibfnamefont{A.}~\bibnamefont{Winter}},
  \bibinfo{journal}{Comm. Math. Phys.} \textbf{\bibinfo{volume}{269}},
  \bibinfo{pages}{107} (\bibinfo{year}{2007}), \eprint{quant-ph/0512247}.

\bibitem[{\citenamefont{Hayden et~al.}(2008)\citenamefont{Hayden, Horodecki,
  Winter, and Yard}}]{HaydenHWY2008-Shannon}
\bibinfo{author}{\bibfnamefont{P.}~\bibnamefont{Hayden}},
  \bibinfo{author}{\bibfnamefont{M.}~\bibnamefont{Horodecki}},
  \bibinfo{author}{\bibfnamefont{A.}~\bibnamefont{Winter}}, \bibnamefont{and}
  \bibinfo{author}{\bibfnamefont{J.}~\bibnamefont{Yard}},
  \bibinfo{journal}{Open Syst. Inf. Dyn.} \textbf{\bibinfo{volume}{15}},
  \bibinfo{pages}{7} (\bibinfo{year}{2008}), \eprint{arXiv:quant-ph/0702005}.

\bibitem[{maj()}]{maj-withrenato}
\bibinfo{note}{We thank Renato Renner for a protocol which achieves this.}

\bibitem[{\citenamefont{Fannes}(1973)}]{Fannes1973}
\bibinfo{author}{\bibfnamefont{M.}~\bibnamefont{Fannes}},
  \bibinfo{journal}{Comm. Math. Phys.} \textbf{\bibinfo{volume}{31}},
  \bibinfo{pages}{291} (\bibinfo{year}{1973}).

\bibitem[{\citenamefont{Alicki and Fannes}(2004)}]{Alicki-Fannes}
\bibinfo{author}{\bibfnamefont{R.}~\bibnamefont{Alicki}} \bibnamefont{and}
  \bibinfo{author}{\bibfnamefont{M.}~\bibnamefont{Fannes}},
  \bibinfo{journal}{J. Phys. A: Math. Gen} \textbf{\bibinfo{volume}{37}},
  \bibinfo{pages}{L55} (\bibinfo{year}{2004}), \eprint{quant-ph/0312081}.

\bibitem[{\citenamefont{Bennett et~al.}(1996)\citenamefont{Bennett, Bernstein,
  Popescu, and Schumacher}}]{BBPS1996}
\bibinfo{author}{\bibfnamefont{C.~H.} \bibnamefont{Bennett}},
  \bibinfo{author}{\bibfnamefont{H.~J.} \bibnamefont{Bernstein}},
  \bibinfo{author}{\bibfnamefont{S.}~\bibnamefont{Popescu}}, \bibnamefont{and}
  \bibinfo{author}{\bibfnamefont{B.}~\bibnamefont{Schumacher}},
  \bibinfo{journal}{Phys. Rev. A} \textbf{\bibinfo{volume}{53}},
  \bibinfo{pages}{2046} (\bibinfo{year}{1996}), \eprint{quant-ph/9511030}.

\bibitem[{\citenamefont{Werner}(1989)}]{Werner1989}
\bibinfo{author}{\bibfnamefont{R.~F.} \bibnamefont{Werner}},
  \bibinfo{journal}{Phys. Rev. A} \textbf{\bibinfo{volume}{40}},
  \bibinfo{pages}{4277} (\bibinfo{year}{1989}).

\bibitem[{\citenamefont{Horodecki and Horodecki}(1999)}]{reduction}
\bibinfo{author}{\bibfnamefont{M.}~\bibnamefont{Horodecki}} \bibnamefont{and}
  \bibinfo{author}{\bibfnamefont{P.}~\bibnamefont{Horodecki}},
  \bibinfo{journal}{Phys. Rev. A} \textbf{\bibinfo{volume}{59}},
  \bibinfo{pages}{4206} (\bibinfo{year}{1999}), \eprint{quant-ph/9708015}.

\bibitem[{\citenamefont{Devetak and Yard}(2006)}]{DevetakY2006-cond-info}
\bibinfo{author}{\bibfnamefont{I.}~\bibnamefont{Devetak}} \bibnamefont{and}
  \bibinfo{author}{\bibfnamefont{J.}~\bibnamefont{Yard}}
  (\bibinfo{year}{2006}), \eprint{quant-ph/0612050}.

\bibitem[{\citenamefont{Oppenheim}(2008)}]{Jono-coherent-relay}
\bibinfo{author}{\bibfnamefont{J.}~\bibnamefont{Oppenheim}}
  (\bibinfo{year}{2008}), \eprint{arXiv:0805.1065}.

\bibitem[{\citenamefont{Schumacher}(1995)}]{Schumacher1995}
\bibinfo{author}{\bibfnamefont{B.}~\bibnamefont{Schumacher}},
  \bibinfo{journal}{Phys. Rev. A} \textbf{\bibinfo{volume}{51}},
  \bibinfo{pages}{2738} (\bibinfo{year}{1995}).

\bibitem[{\citenamefont{Slepian and Wolf}(1971)}]{slepian-wolf}
\bibinfo{author}{\bibfnamefont{D.}~\bibnamefont{Slepian}} \bibnamefont{and}
  \bibinfo{author}{\bibfnamefont{J.}~\bibnamefont{Wolf}},
  \bibinfo{journal}{IEEE Trans. Inf. Theory} \textbf{\bibinfo{volume}{19}},
  \bibinfo{pages}{461} (\bibinfo{year}{1971}).

\bibitem[{\citenamefont{Koashi and Imoto}(2001)}]{KoashiI-compr}
\bibinfo{author}{\bibfnamefont{M.}~\bibnamefont{Koashi}} \bibnamefont{and}
  \bibinfo{author}{\bibfnamefont{N.}~\bibnamefont{Imoto}},
  \bibinfo{journal}{Phys. Rev. Lett.} \textbf{\bibinfo{volume}{87}},
  \bibinfo{pages}{017902} (\bibinfo{year}{2001}), \eprint{quant-ph/0104001}.

\bibitem[{\citenamefont{Christandl et~al.}(2007)\citenamefont{Christandl,
  Ekert, Horodecki, Horodecki, Oppenheim, and Renner}}]{EkertCHHOR2006-ABEkey}
\bibinfo{author}{\bibfnamefont{M.}~\bibnamefont{Christandl}},
  \bibinfo{author}{\bibfnamefont{A.}~\bibnamefont{Ekert}},
  \bibinfo{author}{\bibfnamefont{M.}~\bibnamefont{Horodecki}},
  \bibinfo{author}{\bibfnamefont{P.}~\bibnamefont{Horodecki}},
  \bibinfo{author}{\bibfnamefont{J.}~\bibnamefont{Oppenheim}},
  \bibnamefont{and} \bibinfo{author}{\bibfnamefont{R.}~\bibnamefont{Renner}},
  in \emph{\bibinfo{booktitle}{Proceedings of the 4th Theory of Cryptography
  Conference}} (\bibinfo{publisher}{Lecture Notes in Computer Science},
  \bibinfo{year}{2007}), vol. \bibinfo{volume}{4392}, pp.
  \bibinfo{pages}{456--478}, \eprint{quant-ph/0608199}.

\end{thebibliography}

\end{document}